\documentclass[aps,prc,reprint, twocolumn]{revtex4-1}
\pdfoutput=1
\usepackage{graphicx}				
\usepackage{amssymb}
\usepackage{amsmath}
\usepackage{float}
\usepackage{xcolor}

\begin{document}

\title{One-body Langevin dynamics in heavy-ion collisions at intermediate energies}

\author{Hao Lin and Pawel Danielewicz}
\affiliation{National Superconducting Cyclotron Laboratory and Department of Physics and Astronomy, \\
Michigan State University, East Lansing, Michigan 48824, USA}

\begin{abstract}
We present a new framework to treat the dissipation and fluctuation dynamics associated with nucleon-nucleon scattering in heavy-ion collisions. The two-body collision processes are effectively described in terms of the diffusion of nucleons in viscous nuclear media, governed by a set of Langevin equations in momentum space. The new framework combined with the usual mean field dynamics can be used to simulate heavy-ion collisions at intermediate energies. As a proof of principle, we simulate Au + Au reactions and obtain results consistent with other existing codes under the same constrained conditions. We also study the formation of fragments in Sn + Sn reactions at 50 MeV/nucleon, and results are discussed and compared with two other models commonly employed for collisions.
\end{abstract}

\maketitle

\section{Introduction}

    In heavy-ion collisions at modest energies, two nuclei approach and collide to form a composite nuclear system. At very low incident energies and moderate charges, the system tends to remain fused and de-excites by emission of a few nucleons and light clusters. The picture gains in complexity as the incident energy increases, since more energy is available for the system to populate a greater volume of phase space leading up to a plethora of exit channels. In violent collisions, the composite nuclear system formed is highly excited, and its evolution can be envisioned to be very sensitive to the instabilities present in the system. These instabilities may deform the shape of the system in phase space in an exotic manner, resulting in a breakup into multiple fragments.
    
    The fragmentation phenomenon was experimentally observed \cite{exp1} as early as in the late 1970s. Since the early 1990s, more experimental efforts have been devoted to the study of intermediate-mass-fragment multiplicities \cite{exp2}. 
    
    On the other hand, while different transport models have been successfully applied to describe many one-body observables, our understanding and treatments of the fragmentation mechanism has yet to be unified. The inclusion of fluctuations into transport theories is expected to be of particular importance. It is worth noting that heavy-ion collision experiments make observations and measurements over an \emph{ensemble} of nearly identically prepared colliding systems, and that experimental observables reflect the distribution of all possible outcomes of that ensemble. Angular cross sections, for example, are directly obtained from the angular distribution of the deflected particles, but not from any individual, isolated event. On the contrary, a vast number of the semi-classical transport models are deterministic in nature. These models predict a single exit channel in principle. This poses little trouble when the underlying distribution of outcomes is sharp and narrow. Ensemble averages from transport calculations also prove to converge very well to one-body observables with only weak dependence on channels, such as those concerning collective flows \cite{BauerBertsch,PhysRevLett.54.289}. However, the configurations in multifragmentation are obviously heavily dependent on the exit channels, and, in fact, on the intermediate channels as well. It requires the transport models to be able to explore a wide range of dynamical trajectories. The inclusion of fluctuations creates branching points in the evolution of the system allowing it to jump among different states including those susceptible to instabilities.
    
    In general, there are two major types of transport models for simulating heavy-ion collisions. One type of approaches are essentially molecular dynamics of nucleons represented by single-particle wave-packets, augmented by a phenomenological two-body collision term of the wave-packets \cite{AMD1, *AMD2, IQMD-BNU1, *IQMD-BNU2, *IQMD-BNU3, IQMD, CoMD1, *CoMD2, ImQMD1, *ImQMD2, *ImQMD3, IQMD-IMP1, *IQMD-IMP2, IQMD-SINAP1, *IQMD-SINAP2, *IQMD-SINAP3, TuQMD1, *TuQMD2, *TuQMD3, *TuQMD4, UrQMD1, *UrQMD2}. The propagation and scattering of localized wave-packets help preserve the many-body correlations, and the stochastic treatment of the two-body scattering introduces fluctuations. The other type of approaches aim at directly solving the Boltzmann-Uehling-Uhlenbeck (BUU) equation, with the system characterized by a one-body phase space distribution function \cite{BLOB1, *BLOB2, Gi1, *Gi2, *Gi3, IBL1, *IBL2, *IBL3, pBUU, IBUU, RBUU1, *RBUU2, *RBUU3, RVUU1, *RVUU2, *RVUU3, SMF1, *SMF2, *SMF3}. Solving the BUU equation yields the deterministic time evolution of the one-body distribution function leading to a single exit channel. In recognition of the importance of fluctuations, many efforts have been made to extend the Boltzmann framework, such as the derivation of the Boltzmann-Langevin equation by Ayik and Gregoire \cite{BL1, *BL2}, the Stochastic Mean Field (SMF) model by Colonna \textit{et al.} \cite{SMF1}, and recently the Boltzmann-Langevin-One-Body (BLOB) dynamics by Napolitani and Colonna \cite{BLOB1}. Meanwhile, the inclusion of 3-body scatterings in the pBUU model by Danielewicz \cite{3body} also describes the production of light clusters with mass A $<$ 4.

    In the present article, we propose a simultaneous and consistent treatment of the dissipation and fluctuation in heavy-ion collisions, rather than a mere \emph{ad-hoc} inclusion of fluctuations. We recast the effects of two-body collisions in terms of one-body diffusion processes. This is achieved by replacing the collision integral in the Boltzmann equation by a set of Langevin equations, which govern the seemingly random motion of particles in momentum space. The description of the \emph{beyond-mean-field} dynamics, i.e., the beyond-Vlasov dynamics, is analogous to the classical Brownian motion as we know it, and hence we are tempted to name the new model after Brownian motion. We present the formulation of the theory and describe at length the implementation details in Sec. \ref{II}. In Sec. \ref{III}, we demonstrate the applicability and potential of our model by simulating two different types of heavy-ion collisions, with the latter one focusing on fragmentation dynamics, and we compare our results with other transport models. In the end, a summary is given in Sec. \ref{IV}.

\section{Formulation of the model}\label{II}

    In this section, we will explain the formulation of the Brownian motion model and discuss the details of the implementation of the simulation code.

\subsection{The Boltzmann framework}
    
    In semi-classical transport theories \cite{guide}, the nuclear system is often characterized by the one-body phase space distribution function $f(\mathbf{r}, \mathbf{p}, t)$. The time evolution of the distribution function $f$ is approximated with the Boltzmann equation,
\begin{equation}
\label{Boltzmann}
\frac{\partial f}{\partial t} + \{f, \mathcal{H}\} = I_{coll}.
\end{equation}

    The self-consistent Hamiltonian $\mathcal{H}$ encompasses all information about the nuclear mean field interaction as well as Coulomb interaction, while the residual two-body interaction, mainly nucleon-nucleon scattering, enters through the collision integral $I_{coll}$. The Boltzmann equation provides us with a simple deterministic model to study heavy-ion collisions theoretically. Numerical simulations under the Boltzmann framework can be carried out by means of the test-particle method \cite{guide, LatticeHamiltonian}.

\subsection{The mean-field dynamics}
    
    Neglecting the collision integral in the Boltzmann equation (\ref{Boltzmann}), we recover the so-called Vlasov equation,
\begin{equation}
\label{Vlasov}
\frac{\partial f}{\partial t} + \{f, \mathcal{H}\} = \frac{\partial f}{\partial t} + \frac{\mathbf{p}}{m}\cdot\nabla_{\mathbf{r}}f - \nabla_{\mathbf{r}}U\cdot\nabla_{\mathbf{p}}f = 0,
\end{equation}    
where $\mathcal{H} = \mathcal{T} + \mathcal{V}$ and the mean field $U = \delta \mathcal{V}/\delta \rho$.

    The Vlasov equation retains only one-body information. The interaction between any individual particle and the rest of the system is approximated by a mean-field interaction. In practical calculations, the phenomenological mean-field interactions are usually used,
\begin{eqnarray}
\label{MF}
U_{n/p}\big(\rho(\mathbf{r}), \delta(\mathbf{r})\big) = A \bigg(\frac{\rho(\mathbf{r})}{\rho_0}\bigg) + B \bigg(\frac{\rho(\mathbf{r})}{\rho_0}\bigg)^D \nonumber \\
+ \frac{C}{\rho_0^{2/3}}\nabla^2\bigg(\frac{\rho(\mathbf{r})}{\rho_0}\bigg) \pm 2 S_\textrm{iso}\bigg(\frac{\rho(\mathbf{r})}{\rho_0}\bigg) \delta(\mathbf{r}),
\end{eqnarray}
where $\delta$ = $(\rho_n-\rho_p)/\rho$ is the isospin asymmetry and parameters $A,\ B,\ C,\ D$, and $S_\textrm{iso}$ are fitted to reproduce nuclear matter properties at normal density $\rho_0 = 0.16$ fm$^{-3}$ \cite{LatticeHamiltonian}. Spin dependence and momentum dependence is ignored for simplicity in this parametrization.

    The Coulomb potential $U_\textrm{Coul} (\rho_\textrm{ch}(\mathbf{r}))$ can be determined from the Poisson's equation for electrostatics,
\begin{equation}
\label{Poisson}
\nabla^2 U_\textrm{Coul} = -\frac{1}{\epsilon_0} \rho_\textrm{ch}(\mathbf{r}).
\end{equation}

    In the current model, we consider two species of particles only: neutrons and protons. The numerical scheme of solving the Vlasov equation is adapted from the lattice Hamiltonian method with test particles proposed by Lenk and Pandharipande \cite{LatticeHamiltonian}. The coordinate space is discretized into a cubic lattice with the lattice spacing $l$ = 1 fm. Each test particle has a triangular-shaped form factor and contributes to the nearest eight lattice sites.

\subsection{The dissipation and fluctuation dynamics}
    
    In heavy-ion collisions, nucleon-nucleon scattering acts like a dissipative force, driving the system towards thermal equilibrium. In accordance with the fluctuation-dissipation theorem, the dissipation of the beam energy heats up the system and is thus inevitably accompanied by thermal fluctuations. The thermal fluctuations may manifest themselves in terms of fluctuations in phase space density, which are expected to be linked with multifragmentation observed in intermediate-energy heavy-ion collisions. In this subsection, we aim to develop a framework that offers a consistent and simultaneous description of the dissipation and fluctuation dynamics.
    
    The collision integral tied to two-body scattering reads
\begin{eqnarray}
\label{CollisionIntegral}
I_{coll} = \frac{g}{h^3}\int d^3 p_b \int d\Omega &\,\frac{d\sigma_{ab}}{d\Omega}\, v_{ab} [(1-f_a)(1-f_b)f_{a'}f_{b'} \nonumber \\
&-f_{a}f_{b}(1-f_{a'})(1-f_{b'})],
\end{eqnarray}
where the degeneracy factor is $g = $ 4 for nucleons, and $v_{ab}$ is the relative velocity between nucleons $a$ and $b$, and $d\sigma/d\Omega$ the nucleon-nucleon cross section. The indices $a'$ and $b'$ denote the final states of the colliding pair. 
    
    As the scattering energy increases, nucleon-nucleon scattering peaks more sharply forward. One can reduce the collision integral $I_{coll}$ (\ref{CollisionIntegral}) into a Fokker-Planck form by making an expansion over the scattering angle $\theta$ \cite{FP},
    
\begin{eqnarray}
\label{FP}
I_{coll} \rightarrow -\sum_i \frac{\partial}{\partial p_a^i}\Big\{f_a\cdot\frac{1}{2}\big[\mathbf{\tilde{R}}_a^i+(1-f_a)\mathbf{R}_a^i\big]\Big\} \nonumber \\
+ \sum_{i,j}\frac{\partial^2}{\partial p_a^i \partial p_a^j}\Big(f_a\mathbf{D}_a^{ij}\Big),
\end{eqnarray}
where
\begin{eqnarray}
\label{Coefficients}
\mathbf{R}_a^i &=& -\frac{g}{h^3}\int d^3\,p_b\,f_b\,F_{ab}\,q_{ab}^i,\\
\mathbf{\tilde{R}}_a^i &=& -\frac{g}{h^3}\int d^3 p_b\,f_b\,(1-f_b)\,F_{ab}\,q_{ab}^i, \\
\mathbf{D}_a^{ij} &=& \frac{g}{4h^3}\int d^3 p_b\,f_b(1-f_b)F_{ab}(q_{ab}^2\delta^{ij}\!-q_{ab}^iq_{ab}^j),
\end{eqnarray}
with $\mathbf{q}_{ab} = \mathbf{p}_a - \mathbf{p}_b$ and
\begin{equation}
F_{ab} = (\pi v_{ab}/2) \int_0^1\theta^2\,(d\sigma_{ab}/d\Omega) \,d\cos\theta.
\end{equation}
    
    The \emph{beyond-mean-field} dynamics, depicted traditionally as two-body scattering processes, is transformed, by the Fokker-Planck equation (\ref{FP}), into diffusion processes of nucleons in the viscous nuclear system. The vector coefficients $\mathbf{R}$ and $\mathbf{\tilde{R}}$, usually known as the drag coefficients, are connected to the viscosity of the system. The tensor coefficient $\mathbf{D}$ is referred to as the diffusion matrix, which describes the anisotropic diffusion of particles. 
    
    Note that the differential cross section $d\sigma_{ab}/d\Omega$ enters the equation through the function $F_{ab}$, and hence has an effect on all coefficients in the equation. When an isotropic cross section is used, the next leading order term in the angle expansion yields approximately an extra 13\% contribution to $F_{ab}$ \cite{FP}. One may adjust for the case of non-forward-peaked cross sections by including the higher order terms in the definition of the function $F_{ab}$,
\begin{equation}
F_{ab} = \pi v_{ab} \sum_{k=1}^{\infty} \int_0^1 \frac{\theta^{2k}}{(2k)!} \frac{d\sigma_{ab}}{d\Omega} d\cos\theta.
\end{equation}
In practice, only the second and fourth order terms are included.
    
    Consider an ensemble of systems with identical initial conditions. In the presence of fluctuations, the evolution of the ensemble will diverge. The Fokker-Planck equation provides a mathematical description of the distribution and ensemble-averaged behavior of these identically prepared systems. Indeed, the Fokker-Planck approach has been employed to study the motion of an ensemble of Brownian particles, classical or quantal, in a medium at constant temperature, as the stationary solution of the Fokker-Planck equation yields the equilibrium distribution representing the correct statistics \cite{BALAZS}.
    
    Inspired by the ideas of Brownian motion in a heat bath, we intend to encapsulate the beyond-mean-field dynamics altogether into the Brownian motion of nucleons in the typically non-equilibrium nuclear medium, through the Fokker-Planck equation (\ref{FP}). While the Fokker-Planck equation is deterministic, one may simulate the different dynamical trajectories of the system by use of the corresponding Langevin equation. The differential form of the nonlinear Langevin equation for nucleons undergoing Brownian motion reads,
\begin{equation}
\label{Langevin}
d\mathbf{p}_a = \frac{1}{2}\big[\mathbf{\tilde{R}} + (1-f_a)\mathbf{R}\big]dt + \boldsymbol{\sigma}d\mathbf{B}_t,
\end{equation}
where $\mathbf{R}$ and $\mathbf{\tilde{R}}$ carry the same definitions and meanings as in equations (6) and (7), $\boldsymbol\sigma$ is a 3$\times$3 matrix such that
\begin{equation}
\label{sig}
\mathbf{D}^{ij} =  \sum_{k} \boldsymbol{\sigma}_{ik} \boldsymbol{\sigma}_{jk}.
\end{equation}
$\mathbf{B}_t$ denotes a Guassian random process with properties
\begin{eqnarray}
\langle d\mathbf{B}_t \rangle & = & \mathbf{0}, \\
\langle d\mathbf{B}_t^{i} d\mathbf{B}_t^j \rangle & = & dt\,\delta_{ij}.
\end{eqnarray}

This equation describes the momentum transfer, or the ``kick'', experienced by a nucleon due to its interaction with the medium within a time interval $\Delta t$. The first term is dissipative and connected to the viscosity of the nuclear medium, while the second term is stochastic and gives rise to the fluctuations in the dynamics. In the limit of thermodynamical equilibrium, coefficients in the Langevin equation can be shown to be related by the equilibrium temperature, in a manner akin to the classical Einstein relation.

    For the time being, we do not distinguish between $nn,\ pp,$ and $np$ scatterings by employing a spin-isospin averaged nucleon-nucleon cross section. We further restrict our attention to elastic scatterings by targeting collisions at energies near or below pion production threshold. Extensions to incorporate elastic and inelastic collisions between different species can be made in the future at little cost by adjusting the degeneracy factor $g$ and adopting the suitable differential cross sections $d\sigma_{ab}/d\Omega$. More care needs to be taken for the change of species though in the case of inelastic scatterings.
    
    The Langevin equation meets our goal of treating the dissipation and fluctuations in the dynamics both consistently and simultaneously. We evolve the system by application of the Langevin equation to every nucleon in the system in addition to the mean-field dynamics. Details of the numerical implementations will be discussed in a subsequent subsection.
    
\subsection{Initialization with the Thomas-Fermi equations}
    
    For nucleons inside a stable nucleus, two-body scatterings are strongly suppressed by the Pauli blocking. Hence, for any given mean-field potential, the initial configuration of nucleons in phase space, ideally, should coincide with the stationary solution to the Vlasov equation. This solution amounts to that to the coupled Thomas-Fermi equations \cite{LatticeHamiltonian},
\begin{eqnarray}
\label{TF}
&U_{n}\big(\rho(r), \delta(r) \big)&\, +\, \frac{\hbar^2}{2m_{n}}k_F^2\big(\rho_{n}(r)\big) = \mu_{n}, \\
&U_{p}\big(\rho(r), \delta(r) \big)& +U_\textrm{Coul}+ \frac{\hbar^2}{2m_{p}}k_F^2\big(\rho_{p}(r)\big) = \mu_{p},
\end{eqnarray}
where $U_{X}(\rho, \delta)$ is the self-consistent mean-field potential as in equation (\ref{MF}), $\mu_{X}$ is known as the chemical potential and $k_F$ is the Fermi momentum. The subscript $X$ denotes the particle species.
    
    The self-consistent mean-field potential used is parametrized through equation (\ref{MF}), which contains density-dependent volume and surface terms. The parameters $A$ = $-$209.2 MeV, $B$ = 156.4 MeV, $C$ = $-$6 MeV, $D$ = 1.35, $S_\textrm{iso}$ = 18 MeV and $\rho_0$ = 0.16 fm$^{-3}$ are chosen to reproduce nuclear matter properties at equilibrium: the binding energy of 16 MeV/nucleon, the incompressibility of 240 MeV and the symmetry energy of 30.3 MeV  at normal density $\rho_0$ \cite{CodeComparison}. 
    
    Owning to the surface term, the Thomas-Fermi equations are second-order ordinary differential equations. They are to be solved with the boundary conditions $\rho_{n/p}(r\!\rightarrow\!\infty) = 0$ and $(d\rho_{n/p}/dr)|_{r=0} = 0$.
One may solve them numerically by employing Ansatzes for $\rho_{n/p}(r)$ and adjusting $\mu_{n/p}$ iteratively \cite{pBUU}.
    
    In this work, we propose a different method to solve the coupled Thomas-Fermi equations. We rewrite the equations by multiplying both sides by density $\rho_{X}$,
\begin{equation}
\label{newTF}
h_X\big(\rho_n(r), \rho_p(r)\big)\rho_X(r) = \mu_X\rho_X(r)
\end{equation} 
with the single-particle hamiltonian $h_X = U_X + \hbar^2k_F^2/2m_X + U_\textrm{Coul}\,\delta_{X,p}$. Equation (\ref{newTF}) has the same structure as the Hartree-Fock equation, prompting us to tackle it as an eigenvalue problem. Using a discretized position basis, we can obtain a matrix representation for $h_X$ and a vector representation for $\rho_X$,
\begin{eqnarray}
h_{ij}^{(X)} &=& \langle r_{i} | \hat{h}_X | r_{j} \rangle = h_X\big(\rho_{n/p}(r_i), \rho_{n/p}(r_j)\big), \\
\rho_{i}^{(X)} &=& \langle r_{i} | \rho_X \rangle = \rho_X(r_i).
\end{eqnarray}
Note that $h_X$ is not diagonal, because the derivatives involved are computed in terms of finite differences in the basis. In equation (\ref{newTF}), $\rho_X$ plays the role of the eigenvector of $h_X$ and $\mu_X$ the eigenvalue. We use a self-consistent iterative method \cite{RingSchuck} to find eigenvectors and eigenvalues for $h_n$ and $h_p$. The pair of eigenvectors $\{\rho_n, \rho_p\}$ in the position basis corresponding to the smallest eigenvalues, i.e., lowest chemical potentials $\{\mu_n,\mu_p\}$, is chosen to generate the fields $h_{n/p}(\rho_n, \rho_p)$ in the next iteration and picked as the actual density profiles in the end.
    
\begin{figure}[h]
\begin{center}
\includegraphics[width=0.475\textwidth]{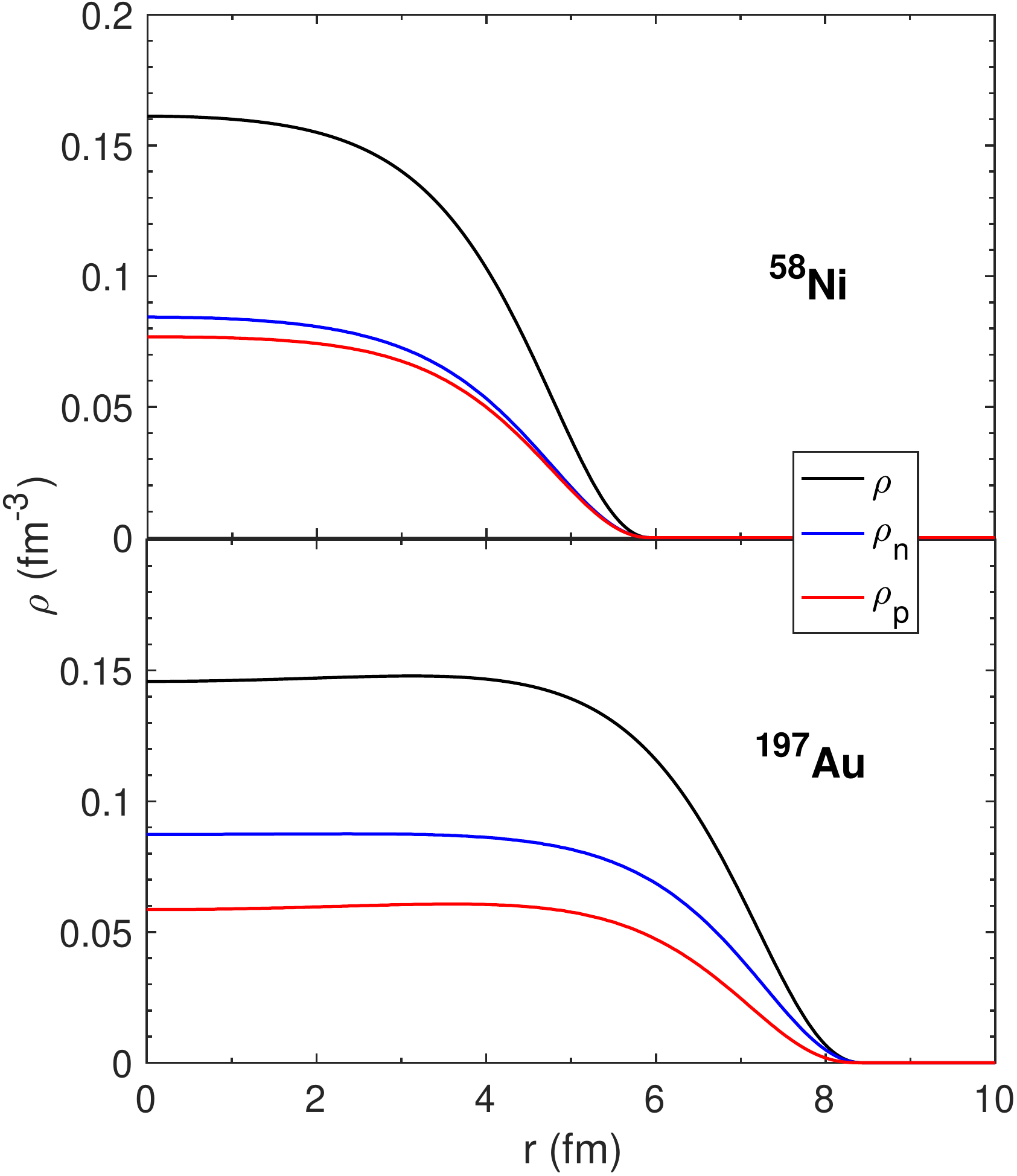}
\caption{Density profiles of stable nuclei $^{58}$Ni and $^{197}$Au obtained from solutions of the Thomas-Fermi equations.}
\label{DP}
\end{center}
\end{figure}

    We demonstrate this method by computing the density profiles for a medium-sized nucleus $^{58}$Ni and a large-sized nucleus $^{197}$Au using the mean-field potential with a $K$ = 240 MeV mentioned above. The computed radial density profiles are shown in Fig. \ref{DP}. We note that density profiles with the Thomas-Fermi approximation lacks the ripples associated with shell effects in typical Hartree-Fock calculations. The tails of the density profiles exhibit rapid fall-offs, which is also typical of Thomas-Fermi calculations \cite{RingSchuck}. In practice, the unphysical fall-off of the tails get mitigated by numerical sampling of test particles and coarse graining, as is seen in the initial distributions at t = 0 fm/c in Fig. \ref{DPTE}. 
    
\subsection{Brownian motions of nucleon wave-packets}
    
    In this model, the beyond-mean-field residual-interactions, between individual nucleons and the nuclear medium they are locally immersed in, are presumed to be dissipative and random, and governed by the proposed Langevin equation (\ref{Langevin}). We refer to these momentum and energy exchanges between particles and media as $\emph{Brownian motions}$. In what follows, we will describe at length the perspective on these Brownian motions and their numerical implementation. Given the mesoscopic nature of nuclear systems, physics and practical details are entangled.

\subsubsection{Partition of test particles into nucleon wave-packets}
    
    While ensuring a sufficiently smooth coverage of the phase space in the simulation of mean-field dynamics, the large quantity of test particles used, typically $N_{test}$ =  $10^2$ - $10^3$ test particles per nucleon, have adverse effects on the fluctuation dynamics. The scatterings of test particles supposedly representing the same nucleon are uncorrelated, which would inevitably wash out most of the fluctuations in the dynamics. This, to a large extent, explains why BUU-type approaches typically have vanishingly small fluctuations compared with QMD-type approaches, whose degrees of freedom are nucleons. Different attempts have been made over the years to restore the nucleonic degrees of freedom in two-body scatterings in the BUU framework \cite{BauerBertsch, BLOB1}.  The main idea of them is to agglomerate test particles adjacent in phase space into so-called nucleon wave-packets and to move them collectively as a whole. 
    
    We adopt a similar approach to enhance the effects of fluctuations. In each time step, we partition the test particles into nucleon wave-packets and execute Brownian motion with these nucleon wave-packets as the degrees of freedom. The pre-partition is accomplished through the k-means clustering algorithm \cite{Statistics} with a metric in phase space parametrized in the following form,
\begin{equation}
\label{metric1}
d^2 = \frac{(\mathbf{r}_i - \mathbf{r}_j)^2}{d_r^2} + \frac{(\mathbf{p}_i - \mathbf{p}_j)^2}{d_p^2}
\end{equation}
where subscripts $i$ and $j$ denote two points in phase space. The parameters $d_r$ and $d_p$ address the compactness in the coordinate and the momentum spaces respectively. We run the k-means clustering algorithm to partition both the neutron test particles and the proton ones separately. The algorithm is set to terminate after several iterations, and the values $d_r$ = 1.2 fm and $d_p$ = 130.5 MeV/c are used. It is found in practice that the final results are not sensitive to either the early termination of the clustering algorithm or the values of the metric parameters. 
    
    After the pre-partition, the system is divided into $N$ neutron subspaces and $Z$ proton subspaces. The pre-partition is simple but somewhat arbitrary, and thus only the centroids are to be used. We identify these centroids as the \emph{scattering centers} for the nucleons. 
    
    For each centroid $(\mathbf{r}_i, \mathbf{p}_i)$ for which the local nucleon density is above 0.1 fm$^{-3}$, we consider a spherical region centered at $\mathbf{r}_i$ of radius $R \sim 2$ fm. This value corresponds roughly to the sum $\sqrt{\sigma_{NN}/\pi} + \sqrt{\langle r_{ch}^2 \rangle}$, where the nucleon-nucleon cross section $\sigma_{NN} \simeq$ 42 mb and the root-mean-square proton charge radius $\sqrt{\langle r_{ch}^2 \rangle} \simeq$ 0.86 fm. $R$ can also be made density-dependent, with the sensible choice of $R(\rho) = [2.01 - 0.18\rho^{-\frac{4}{3}}(\rho-\rho_0)]$fm. Inside the spherical region, we search for test particles close to the centroid $(\mathbf{r}_i, \mathbf{p}_i)$ using the following phase-space metric:
\begin{equation}
\label{metric2}
d^2 = \frac{(\mathbf{r}_i - \mathbf{r}')^2}{\sigma_r^2} + \frac{(\mathbf{p}_i - \mathbf{p}')^2}{\sigma_p^2}
\end{equation}
with $\sigma_r = R$ and $\sigma_p = {\hbar}/2R$. This metric emphasizes the compactness in momentum space, while facilitating the Heisenberg uncertainty principle. The Langevin equation (\ref{Langevin}) is originally intended for point-like particles, and hence wave-packets well-localized in momentum space are preferred. The $N_{test}$ test particles of the same species closest to the centroid form the wave-packet to undergo Brownian motion. The rest of the particles constitute the medium, with which the wave-packet interacts.

\subsubsection{Evaluations of the Langevin equation's coefficients on a lattice}
    
    The coefficients $\mathbf{R}$, $\mathbf{\tilde{R}}$ and $\mathbf{D}$ involve integrals folded over the momentum space, which require the knowledge of occupation at different momenta. To this end, we construct a three-dimensional cubic lattice over the entire momentum space inside each spherical scattering region and evaluate the occupation at different sites.
    
    The lattice spacing $L_p$ needs to be chosen with care to faithfully reflect the actual spread and spacing of the underlying test particles. We use the standard deviation $\sigma_{p}^{wp}$ of the momentum of test particles belonging to the nucleon wave-packet as a measure to constrain lattice spacing. We normally choose the spacing $L_p = \max \{\sqrt[3]{2}\sigma_{p}^{wp}, \hbar/R \}$, where R is the radius of the spherical scattering region under consideration. With such constraints, the values of the spacing $L_p$ typically fall between 100 MeV/c and 140 MeV/c, which ensures a sensible coarse graining of the momentum space. 
    
    Occupation $f(\mathbf{p})$ at each momentum lattice site is evaluated in the same fashion as spatial densities are on a spatial lattice in mean-field dynamics simulations. Test particles have a triangular-shaped form factor, contributing to the eight nearest lattice sites only. Integrals of $\mathbf{R}$, $\mathbf{\tilde{R}}$ and $\mathbf{D}$ are computed as summations over all sites of the three-dimensional momentum lattice. 

\subsubsection{Momentum transfer from the nuclear medium to the nucleon wave-packet}
    
    With a reasonable time step size $\Delta t \sim$ 0.25 - 0.5 fm/c, the first term of the Langevin equation (\ref{Langevin}) can be readily calculated. The stochastic term involves a matrix $\boldsymbol\sigma$, which needs to be extracted as the square root of the diffusion matrix $\mathbf{D}$. Note that, by the definition in equation (8), $\mathbf{D}$ is a real symmetric positive semi-definite matrix. It follows that $\mathbf{D}$ can be diagonalized as $\mathbf{D} = \mathbf{O} \mathbf{\Lambda} \mathbf{O}^\top$, where $\mathbf{\Lambda}$ is a diagonal matrix and $\mathbf{O}$ an orthogonal matrix. We then represent $\boldsymbol\sigma$, cf. equation (\ref{sig}), as
\begin{equation}
\label{sigma}
\boldsymbol{\sigma} = \mathbf{O} \mathbf{\Lambda}^{\frac{1}{2}} \mathbf{O}^\top,
\end{equation}
where $\mathbf{\Lambda}^{\frac{1}{2}}$ is diagonal whose diagonal elements are the unique square roots of the corresponding diagonal elements in $\mathbf{\Lambda}$. It can be easily verified that $\boldsymbol\sigma$ constructed according to (\ref{sigma}) satisfies (8).
    
    With a time step of size $\Delta t$, the differential notation $d\mathbf{B}_{t}$ is interpreted as a 3-dimensional random vector, whose components are independent Gaussian random numbers. The underlying Gaussian distribution has a mean equal to zero and a variance equal to $\Delta t$. 
    
    In summary, the momentum transfer $\Delta\mathbf{p}$ within a time step $\Delta t$ in a spherical scattering region is simulated as
\begin{equation}
\label{Langevin2}
\Delta\mathbf{p}_a = \frac{1}{2}\big[\mathbf{\tilde{R}} + (1-f_a)\mathbf{R}\big]\Delta t + \boldsymbol{\sigma}\,\mathbf{g}(0, \Delta t)
\end{equation}
with $\mathbf{g}(0, \Delta t)$ being a random vector comprised of 3 independent Guassian random numbers sampled with mean = $0$ and variance = $\Delta t$.

\subsubsection{Recoil for conservation of momentum and energy}
    
    After a nucleon wave-packet is shifted in the nuclear medium inside a spherical scattering region, the recoil of the nuclear medium needs to be accounted for in order to preserve the conservation of total momentum and total energy. The interaction between the nucleon wave-packet and the nuclear medium is reciprocal. Indeed, by exchanging the subscripts $a$ and $b$ in the Langevin equation (\ref{Langevin}), one obtains an expression for how the nucleon wave-packet induces recoil of particles in the medium. Thus, the recoil can be treated in principle precisely.
    
    On the other hand, owing to the facts that the number of particles involved in the medium is large and that the recoils are coupled in a nontrivial manner, we instead adopt a collective and approximate treatment of the recoil effects: the center of momentum of the nuclear medium is shifted to conserve total momentum, and all particles in the medium are scaled with respect to the new center of momentum to conserve total energy. 
    
    Additionally, it is worth noting that the nuclear medium almost always contains more than one nucleon. The collective shift-and-scale adjustment, in effect, introduces many-body correlations in the nuclear medium.
    
\subsubsection{Pauli-blocking procedure}

\begin{figure}[h]
\begin{center}
\includegraphics[width=0.475\textwidth]{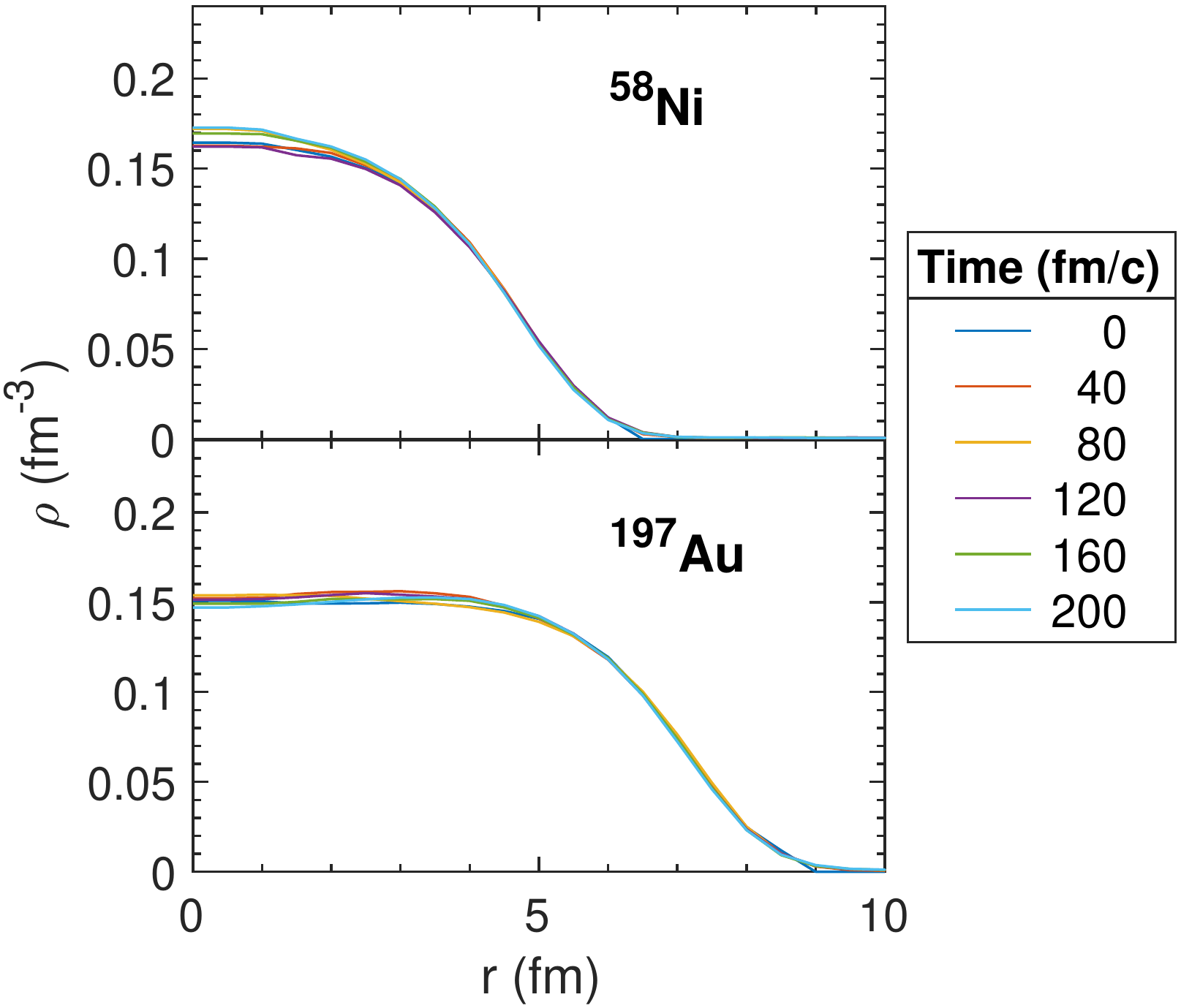}
\caption{Time evolution of radial density profiles for single nuclei $^{58}$Ni and  $^{197}$Au in steps of 40 fm/c.}
\label{DPTE}
\end{center}
\end{figure}

    Within the scattering region, after the kick of the nucleon wave-packet and the adjustment for recoil in the nuclear medium, we compute again the occupation over the entire lattice in momentum space. The Brownian motion is finalized only if none of the occupation at any lattice site exceeds 1. Otherwise, we deem the Brownian motion unphysical and revert all changes. This  Pauli blocking procedure proves to be effective. For single ground state nuclei, over 97\% of the attempted Brownian motions are $\emph{blocked}$ in the current model prescription, and stability of the nuclei are demonstrated in Fig. \ref{DPTE}.
        
Fig. \ref{DPTE} shows the time evolution of the radial density profiles for a single $^{58}$Ni and a single $^{197}$Au up to 200 fm/c in steps of 40 fm/c, simulated by the our Brownian code under the same controlled conditions specified in Ref. \cite{CodeComparison}. Ideally, the density distribution should remain unchanged over time for single nuclei. In our case, the density profiles show only small scale fluctuations over the entire course of simulation. It indicates, in particular, that numerical solutions of the Thomas-Fermi equations (\ref{TF}) approximate the true solutions of the Vlasov equation (\ref{Vlasov}) reasonably well. Further, majority of the spurious large momentum transfers are effectively blocked by the Pauli-blocking procedure.
    
\subsubsection{Summary of implementation of Brownian motions}

    In each time step, the occupied phase space is partitioned into $N$ + $Z$ subspaces of roughly equal volume. Scattering regions are constructed spherically around the spatial centroids of each subspace. These regions are to be examined successively in a random order. Within each scattering region, a separation of the nucleon wave-packet from the nuclear medium is made. The Langevin equation is evaluated on a 3-dimensional lattice in momentum space, and the resulted momentum transfer is applied to the nucleon wave-packet. In observance of the conservation laws, the recoil effects are taken into account through an adjustment of the momentum distribution of particles in the medium. A Pauli-blocking procedure is applied in the end to preserve the Pauli exclusion principle.

\section{Results} \label{III}

In this section, we first demonstrate practical applicability of our model to heavy-ion collisions by simulating the Au-Au collisions at both $100$ MeV/nucleon and $400$ MeV/nucleon at an impact parameter of $b = 7$ fm. We study the nucleon rapidity distribution and the average in-plane flow $\langle p_x/A \rangle$ and compare our simulation results with those from other transport codes in Ref. \cite{CodeComparison}. After confirming that our model can yield reasonable results for one-body observables, we proceed to investigate its ability to describe multi-fragmentation processes. To this end, we study the time evolution of the systems: $^{112}$Sn + $^{112}$Sn and $^{124}$Sn + $^{124}$Sn at $50$ MeV/nucleon with an impact parameter $b = 0.5$ fm. A preliminary comparison of results with the Stochastic Mean Field (SMF) and the Antisymmetrized Molecular Dynamics (AMD) models from Ref. \cite{SMFAMD} is also made.

\subsection{One-body observables for Au + Au system}

Fluctuations associated with Brownian motions enable our model to probe a broader range of intermediate and final channels. It is of interest to study whether the diversity of intermediate and final channels may affect the description of one-body observables. Our model is applied to simulate the Au + Au reactions at two incident energies, 100 MeV/nucleon and 400 MeV/nucleon. These specific reactions were also studied and compared in a transport code comparison project under controlled conditions \cite{CodeComparison}. The same impact parameter $b = $ 7 fm as there is chosen. Identical mean-field interactions and nucleon-nucleon cross sections as there are employed. 

\subsubsection{Rapidity distribution}

The rapidity distribution in the final state gives a Lorentz invariant measure of the degree of stopping of nucleons attained in heavy-ion collisions \cite{CodeComparison}.  The more particles populate the mid-rapidity region in the center of mass frame, the stronger the stopping effects are. 

\begin{figure}[h]
\begin{center}
\includegraphics[width=0.475\textwidth]{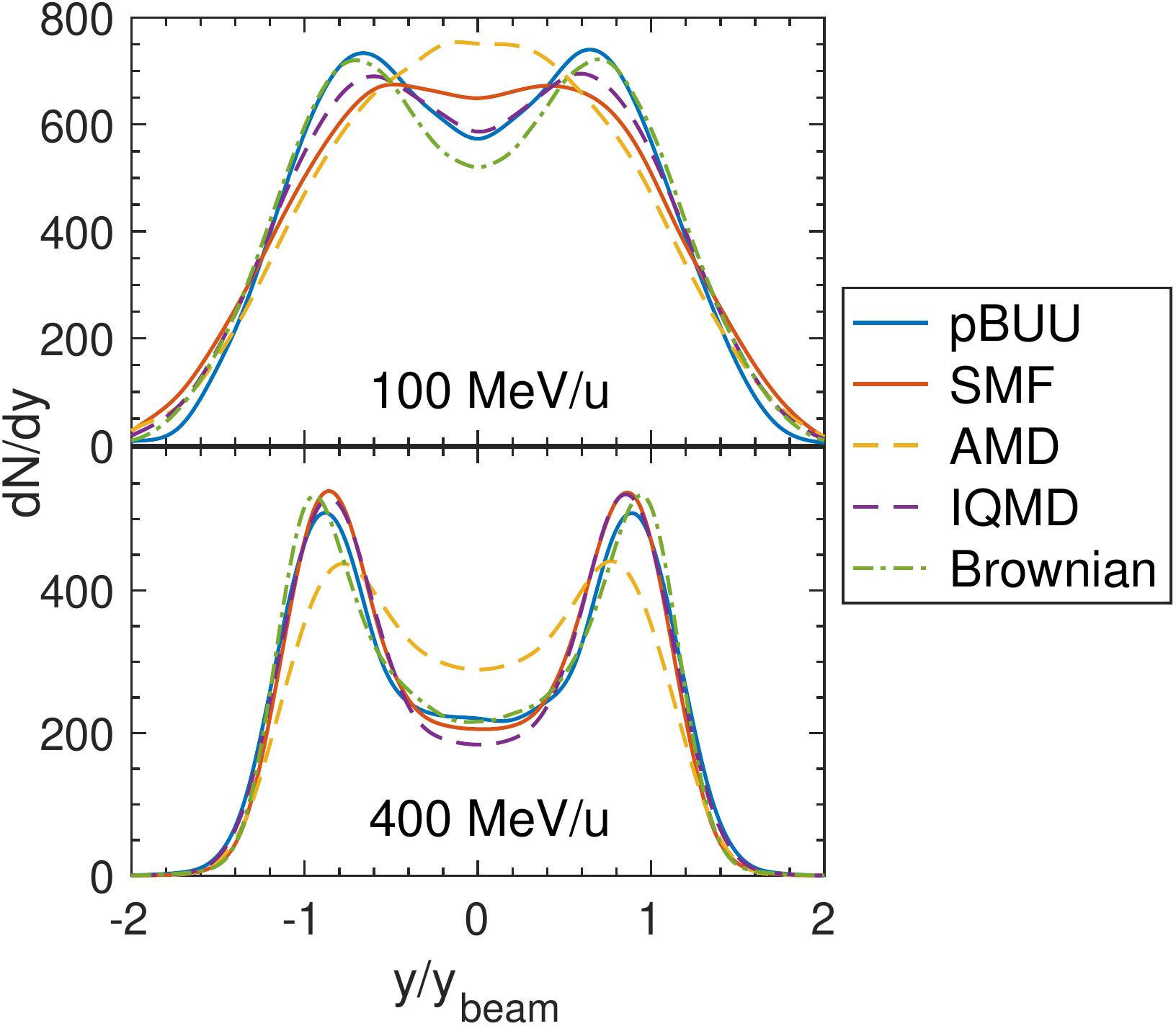}
\caption{Final rapidity distributions as a function of reduced rapidity for $^{197}$Au + $^{197}$Au at beam energies of 100 MeV/nucleon (upper panel) and 400 MeV/nucleon (lower panel) with an impact parameter $b$ = 7 fm. Solid curves, dashed curves and the dashed-dot curve correspond to BUU-type models, QMD-type models \cite{CodeComparison} and the Brownian model, respectively.}
\label{RD}
\end{center}
\end{figure}

In Fig. \ref{RD}, we display the final rapidity distributions from our calculations accompanied by results of selected BUU and QMD calculations from Ref. \cite{CodeComparison}. At low incident energy 100 MeV/nucleon, there is a large amount of filling of the mid-rapidity region, indicating a relatively strong stopping. While all codes except for AMD exhibit a shallow double-humped structure, differences in details of the rapidity distributions are not negligible. This is probably tied to differences in treating Pauli principles in different codes and delicate competition of mean-field interaction and many-body correlations at this incident energy \cite{CodeComparison}. At higher incident energy of 400 MeV/nucleon, fewer particles populate the mid-rapidity region compared to the outer regions and the double-peaked feature is more pronounced, indicating a weaker stopping due to shrinking Fermi momentum compared to incident momentum per nucleon. General consistency is found among most calculations, with AMD as an outlier predicting a stronger stopping.

\subsubsection{Average in-plane flow $\langle p_x/A \rangle$}

Use of finite impact parameter, $b = $ 7 fm, in this study,  breaks the macroscopic rotational symmetry around the beam axis in the system, and therefore anisotropy appears in the transverse collective momentum distribution. We focus on the average in-plane flow $\langle p_x/A \rangle$, simply known as the transverse flow, as a function of the reduced rapidity $y/y_{beam}$ in the center of mass frame. When quantified, the transverse flow is commonly described in terms of an ``S-shaped'' curve. The slope at the origin, commonly known as the ``slope parameter'', is of importance. Particles in the mid-rapidity region are expected to come from the compressed region during the collision, and thus the study of this flow parameter can shed light on the behavior of the nuclear equation of state beyond normal density.

\begin{figure}[h]
\begin{center}
\includegraphics[width=0.475\textwidth]{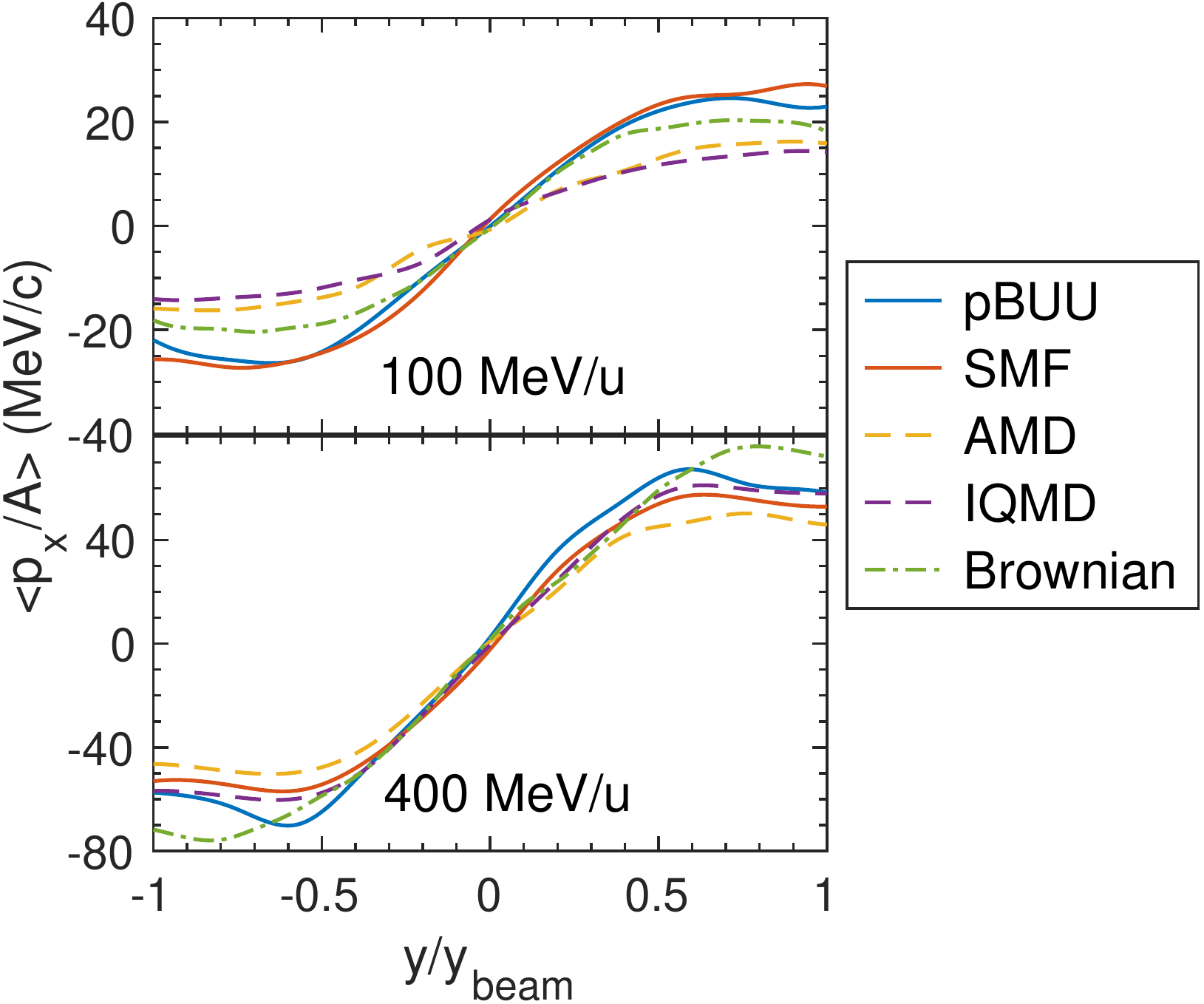}
\caption{Final average in-plane flow as a function of reduced rapidity for $^{197}$Au + $^{197}$Au collisions at beam energies of 100 MeV/nucleon (upper panel) and 400 MeV/nucleon (lower panel) and an impact parameter $b$ = 7 fm. Solid curves, dashed curves and the dashed-dot curve represent BUU-type models, QMD-type models \cite{CodeComparison} and the Brownian model, respectively.}
\label{flow}
\end{center}
\end{figure}

In Fig. \ref{flow} we show the average in-plane flows of our calculations together with results from selected transport models from Ref. \cite{CodeComparison}. In all calculations, the expected S-shaped curves are produced at both energies. The positive slopes at the origin indicate that the effects of nucleon-nucleon scattering dominate over those of the mean field. At 100 MeV/nucleon, the BUU-type models clearly produce greater inflections than the QMD-type models. The prediction from the Brownian motion model lies between them. At 400 MeV/nucleon, it appears that all five transport models yield very consistent results in the mid-rapidity region.

\begin{figure}[h]
\begin{center}
\includegraphics[width=0.475\textwidth]{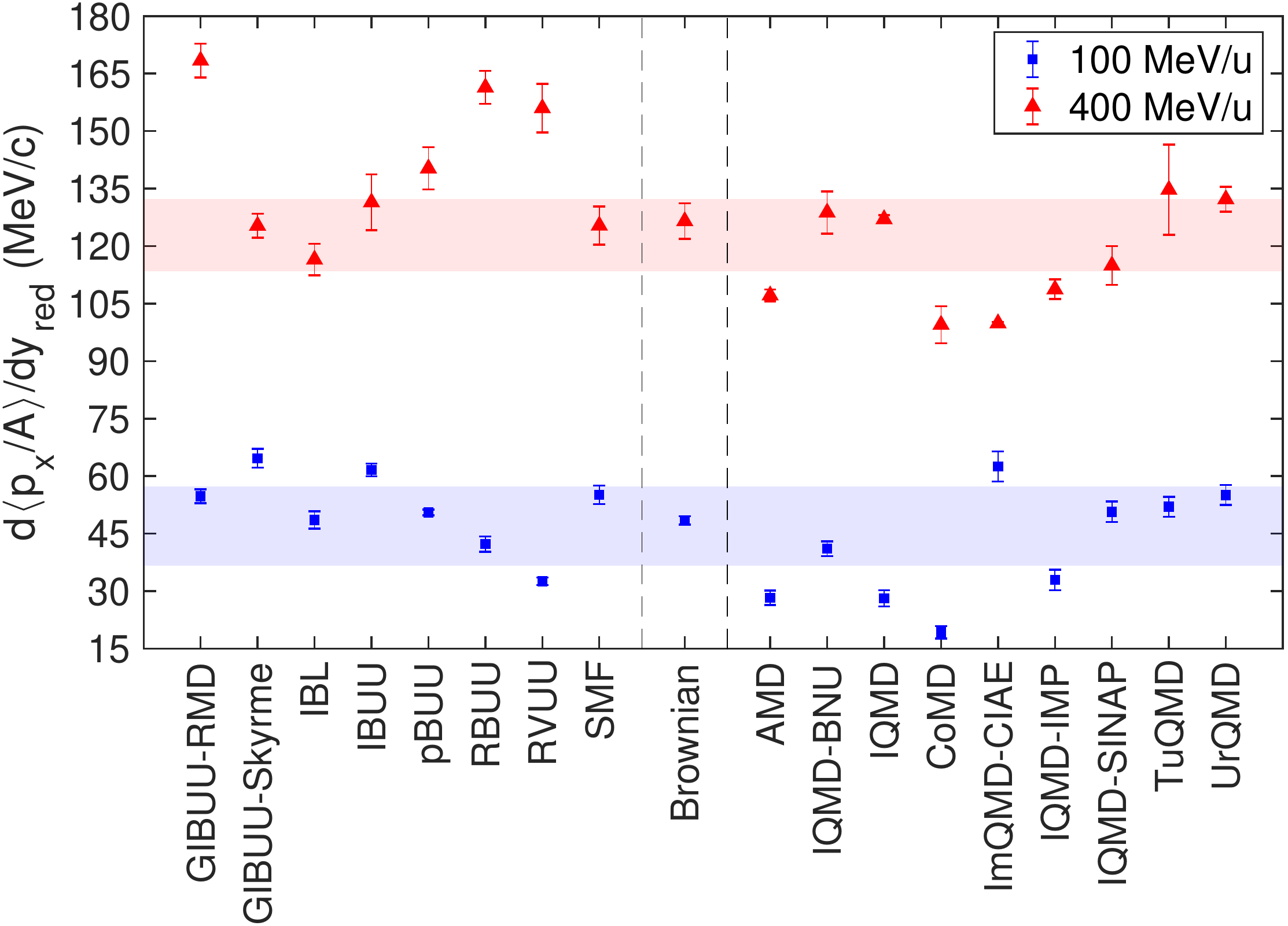}
\caption{Flow slope parameters for different transport models \cite{CodeComparison} for $^{197}$Au + $^{197}$Au collisions at beam energies of 100 MeV/nucleon (blue squares) and 400 MeV/nucleon (red triangles) at an impact parameter $b$ = 7 fm. The error bars represent the fitting uncertainties. These become invisible when they are smaller than the symbols. The colored bands correspond to roughly 52\% confidence intervals from the statistics of calculations from both the BUU-type and the QMD-type models. (See text for a more detailed explanation.)}
\label{slope}
\end{center}
\end{figure}

The slope parameters at mid-rapidity can be extracted through a linear fit in a small interval centered at the origin. The values of the slope parameters at two energies for different transport simulations \cite{CodeComparison} are summarized in Fig. \ref{slope}. The error bars take into account the fitting uncertainties only. On top of the simulation results, we also added shaded bands to indicate regions in which calculations are considered to be \emph{statistically} consistent with majority of the BUU-type and QMD-type models. To obtain the statistically consistent regions, we first computed the $\sqrt{2}\sigma$-intervals centered at the mean values for the BUU sample and the QMD sample independently. $\sigma$ stands for the standard deviation of the sample. The statistically consistent regions were taken to be the overlap of the $\sqrt{2}\sigma$-intervals from the two samples. If we further assume that the BUU sample and the QMD sample follow an identical Gaussian distribution, the consistent regions can also be interpreted as roughly 52\% confidence intervals. Note that throughout the statistical analysis, results from the Brownian motion model were deliberately excluded to avoid any possible bias. Nevertheless, the slope parameters extracted from the Brownian simulations are found to be statistically consistent with majority of the other calculations. 

The reassuring consistency of results between the Brownian motion model and other current transport models provides evidence that the one-body Brownian motion picture can successfully capture the effects of two-body scattering in heavy-ion collisions. 

Last but not least, in analyzing results from the Brownian motion model, we averaged the rapidity and the in-plane flow over 32 independent events, and the averaged results exhibit good parities as functions of the reduced rapidity. It needs to be pointed out that the fluctuation dynamics described by the Langevin equation (\ref{Langevin}) does not automatically preserve forward-backward reflection symmetry, and that parity symmetry breaking or other types of symmetry breaking can be observed in individual events. Since the ``directions'' of the symmetry breakings are essentially random, symmetry can be restored by ensemble-averaging over independent runs. In fact, these symmetry breakings, resulted from the broadening of dynamical trajectories, can be crucial for the formation of fragments to be discussed in the next subsection. For instance, an uneven break-up of two nuclei in symmetric central collisions will be likely associated with an asymmetry in the rapidity distribution.

\begin{figure*}
\begin{center}
\includegraphics[width=0.95\textwidth]{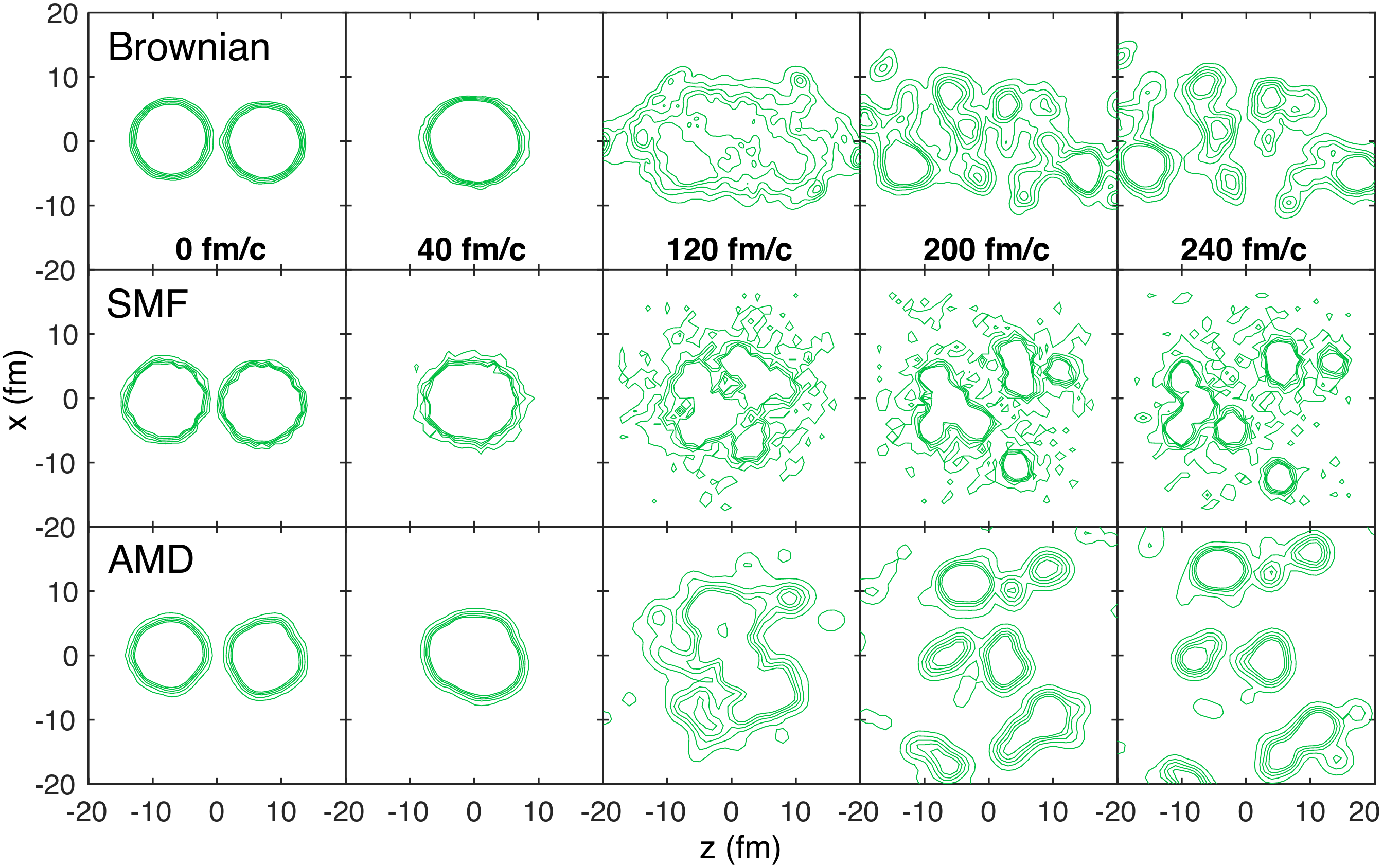}
\caption{Time evolution of density contours for nucleons projected onto the reaction plane in the reaction $^{112}$Sn + $^{112}$Sn at 50 MeV/nucleon at an impact parameter of b = 0.5 fm, at several different times. Results calculated with the Brownian motion model are displayed in the first row of the panels. The bottom two rows show results from SMF and AMD calculations \cite{SMFAMD}. The densities of the projected contours begin at 0.07 fm$^{-2}$ and increase at intervals of 0.1 fm$^{-2}$.  }
\label{contour}
\end{center}
\end{figure*}

\subsection{Fragmentation dynamics with Sn + Sn at 50 MeV/nucleon}

In this subsection, we will demonstrate how the ability of our model to probe a plethora of intermediate and final states is connected with the formation of intermediate mass fragments (IMF) in central heavy-ion collisions. The systems we study are $^{112}$Sn + $^{112}$Sn and $^{124}$Sn + $^{124}$Sn at 50 MeV/nucleon at an impact parameter b = 0.5 fm. The evolution of the systems is followed up to 280 fm/c after initial contact. These systems have already been studied by Colonna \textit{et al.} using both SMF and AMD \cite{SMFAMD}. We will make a preliminary comparison between our simulation and those in SMF and AMD.

Fig. \ref{contour} shows the density contour plots from projecting nucleons on the reaction plane for the $^{112}$Sn + $^{112}$Sn system at different stages of the reaction calculated with different transport models. All three models give a qualitatively consistent description of the compression-expansion dynamics, and cluster structures are formed in the expansion phase. During the approach and compression up to 40 fm/c, calculations from the three models with fluctuations and many-body correlation effects do not appear to be distinguishable from what would be expected from conventional transport models without fluctuations. The suppression in the role of fluctuations is linked to the limited volume of phase space available for the system to populate, since it is far from thermalized in the early stage. At 120 fm/c, we can already observe, in all simulations, that the expanding systems turn inhomogeneous. These inhomogeneities, which would not have existed without fluctuations, provide seeds for fragmentation, and there are cluster structures forming in the core. As the system continues to expand, the lumps of matter move away from one another and escape from the central region. For all three models, sizable fragments can be identified after 200 fm/c, and changes between 200 fm/c and 240 fm/c are sporadic and moderate. As a result, we assume that the configurations have frozen out by 240 fm/c, and we terminate the simulations at 280 fm/c. 

The three models differ quite substantially in details of the predictions for the expansion phase. In the Brownian motion model calculations, the degree of stopping is comparatively low and the system tends to expand more along the beam direction. On the other hand, in AMD, the system expands quickly with a focus around the \emph{x}-direction, which indicates a very strong stopping that is also seen in the Au-Au simulations. The relatively isotropic and slow expansion in SMF can probably be explained by the spinodal decomposition of a nearly homogeneous source at low density \cite{spinodal}. We also count the number of nucleons in the ``gas'' phase ($\rho < \frac{1}{6} \rho_0$) predicted by our model and compare the number to results of SMF and AMD \cite{SMFAMD2007}. It is found that our model yields more gas-phase nucleons than AMD, but only slightly fewer than SMF. 

Regarding the fragmentation mechanism in our model, in-medium Brownian motions introduce branching points in the dynamical trajectories and allow for ``jumps'' among a greater range of intermediate and final configurations. These jumps, stemming from nucleon-nucleon scatterings, are abrupt and discontinuous in time. The Langevin equation (\ref{Langevin}), with its stochastic term introducing discontinuities in time, serves the purpose. By solving the Langevin equations, we attempt to simulate the abrupt jumps from one n-body configuration to another.  As the deviations from the ensemble-averaged trajectory predicted by the Boltzmann equation (\ref{Boltzmann}) accrue from the jumps, exotic configurations including those with fragmentation eventually become accessible. In the quantum-mechanical picture, configurations are represented by superpositions of Slater determinants. While mean-field evolution is coherent in principle, the residual incoherent many-body correlations, such as two-body scattering, result in decoherence and transitions between different Slater determinants. Among the stochastic approximations of the quantum many-body problems are the AMD model \cite{AMD1, *AMD2, AMD3} and the Stochastic Time-Dependent Hartree-Fock (STDHF) theory \cite{STDHF}. In treating the jumps between different configurations as a stochastic process, the Brownian motion model is conceptually consistent with these quantal approaches.

\begin{figure}[h]
\begin{center}
\includegraphics[width=0.475\textwidth]{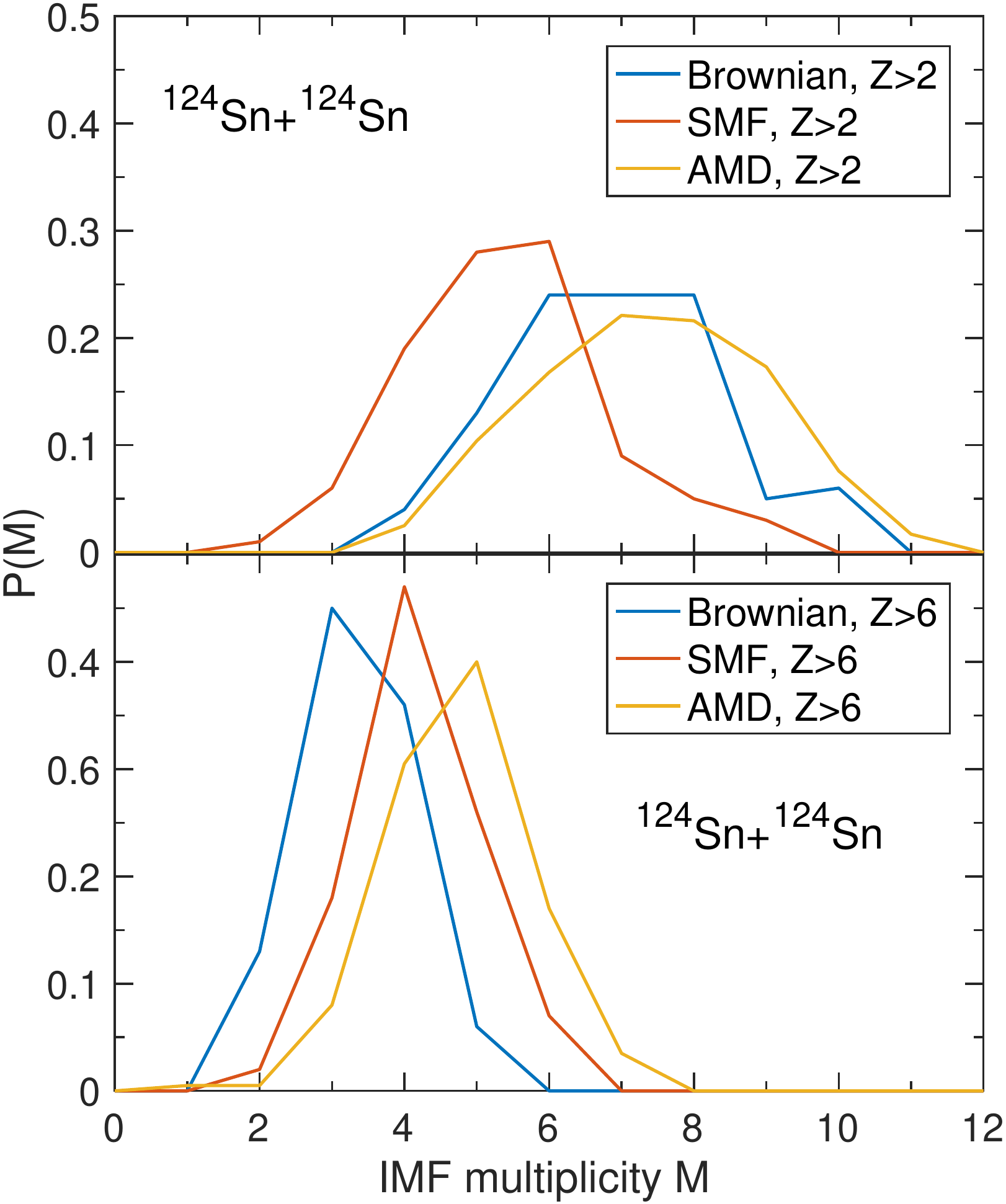}
\caption{Distribution of IMF multiplicity obtained from different models for the reaction $^{124}$Sn + $^{124}$Sn at 50 MeV/nucleon with an impact parameter b = 0.5 fm. The upper panel shows the multiplicity distribution of IMFs with charge Z $>$ 2 and the lower panel with Z $>$ 6 \cite{SMFAMD}. }
\label{multiplicity}
\end{center}
\end{figure}

A comparison of the IMF multiplicity between the Brownian motion and the other two transport models for $^{124}$Sn + $^{124}$Sn is shown in Fig. \ref{multiplicity}. In our case, fragments are identified with a simple coalescence algorithm with a cut-off density $\rho_c$ = 0.02 fm$^{-3}$ \cite{BauerBertsch}. The distribution is obtained from 100 independent simulations.

For IMFs with charge Z $>$ 2, our multiplicity distribution looks compatible with that from AMD. Both distributions maximize around multiplicity $=$ 7 and share a similar spread. For larger IMFs with charge Z $>$ 6, our calculations yield a slightly lower multiplicity, while still predicting essentially the same spread as the others. Since our model, as well as SMF, predicts more free nucleon emission than AMD, fewer nucleons are available in the ``liquid'' phase ($\rho > \frac{1}{6} \rho_0$) for the assembly of IMFs. Moreover, due to the large number of free nucleons and light fragments in our simulations, the effective surface-to-volume ratio is also higher, which might lead to spurious evaporation of test particles and hinder the formation of fragments. Nonetheless, the Brownian motion model breaks through the limitations of traditional Boltzmann transport framework and proves to have great potential for the description of multifragmentation.

\section{Summary and discussion} \label{IV}

In this paper, we have reformulated the beyond-mean-field dynamics in heavy-ion collisions in terms of Brownian motions of nucleons in the viscous, out-of-equilibrium nuclear medium, as opposed to the typical two-body scatterings. The Brownian motions are, in effect, the momentum and energy exchange between a nucleon and the nuclear medium it is immersed in. They are governed by a set of Langevin equations consisting of a friction-like term and a stochastic term. This approach describes the dissipation and fluctuation dynamics consistently and simultaneously. Furthermore, each simulation generates a unique dynamical trajectory, enabling us to probe different exit channels and obtain the distribution of possible outcomes of the ensemble. 

The details of the numerical simulations, including a new method to initialize stable nuclei from the Thomas-Fermi approximation, have been presented. We have applied our model to the time evolution of isolated stable nuclei. The stabilities of the simulations and the nuclei are well established. 

To demonstrate that our model's ability to describe one-body observables is on par with that of other transport models, we have studied the final rapidity distribution and average in-plane flow in the reaction $^{197}$Au + $^{197}$Au at two incident energies and showed that our results are comparable with those obtained from either QMD-type models or BUU-type models \cite{CodeComparison}. 

We have also investigated formation of fragments in heavy-ion collisions with our model and confirmed the crucial role fluctuations play in seeding multifragmentation. We have repeated the calculations of Sn + Sn at E = 50 MeV/nucleon, previously done with the SMF and the AMD models \cite{SMFAMD}. As seen from the time evolution of density contours for nucleons projected on the reaction plane, and all three models depict a fragmented system with similar general features. Regarding the distribution of IMF multiplicity, We find that the yield of light IMFs with Z $>$ 2 in the Brownian motion model is comparable to that in AMD, but our yield of large IMFs with Z $>$ 6 is slightly lower than that in SMF or AMD.

So far, we have successfully demonstrated the abilities and potential of the Brownian motion model to describe various scenarios in heavy-ion collisions at intermediate energies. Its ability to traverse different dynamical trajectories makes it particularly suitable for the study of multifragmentation, which is beyond the reach of many traditional transport models. It is also superior to many models with stochastic extensions in that it treats dissipation and fluctuation on an equal footing. More explicit introductions of many-body correlations are under consideration. In the future, it is of great interest to confront the optimized Brownian motion model to experimental data. We have also in mind the goal of studying the fragmentation mechanism. For example, we have looked at whether and when the central region of a fragmented system enters the mechanically unstable region and the findings seem to favor the spinodal decomposition mechanism. A more careful inspection is planned.

\begin{acknowledgements}
This work was supported by the National Science Foundation, under Grant No. PHY-1403906 and PHY-1520971.

\end{acknowledgements}

\bibliography{Brownian.bib}

\end{document}